\documentclass[10pt,journal,final]{IEEEtran}
\usepackage{graphicx}
\usepackage{hyperref}
\begin{document}

\title{Structure of magnetic domain wall in cylindrical microwire}
\author{\IEEEauthorblockN{Andrzej Janutka\IEEEauthorrefmark{1} and Przemys{\l}aw Gawro\'nski\IEEEauthorrefmark{2}}\\
\IEEEauthorblockA{\IEEEauthorrefmark{1}Institute of Physics, Wroclaw University of Technology, 50-370 Wroc{\l}aw, Poland\\
\IEEEauthorrefmark{2}Faculty of Physics and Applied Computer Science, AGH University of Science and Technology, 30-059 Krakow, Poland}
\thanks{E-mail: Andrzej.Janutka@pwr.wroc.pl}}
\maketitle

\begin{abstract}
Within a simple model, we study magnetic domain walls (DWs) inside the inner core
 of the amorphous ferromagnetic microwire whose spin ordering is a core-shell structure.
 The interaction of the (internal-stress created) outer shell of the wire on the inner core
 is included into the Landau-Lifshitz equation via an effective Dzyaloshinskii-Moriya-like
 anisotropy. Resulting DW textures are classified. The model is applicable to the 
 nanowires of a modulated diameter (periodically constricted nanowires) as well.
 In that case a core-shell magnetic structure is of
 the purely-magnetostatic origin. Because the micromagnetic simulations of the microwires
 are extremely challenging, the simulations of the structured nanowires are performed
 with the purpose of verifying analytical predictions on the shape of the DW.
\end{abstract}


\maketitle

\section{Introduction}

Magnetic amorphous glass-coated wires of the micron-sized diameter are very important 
 objects for sensor manufacturing while the process of the wire remagnetization 
 [via the domain-wall (DW) propagation] is a basics of the sensing mechanism
 (next to the giant magneto-impedance; GMI effect). However, the sufficiently detailed while
 simple model of the DW statics and dynamics that would enable the description 
 of the microwire remagnetization is still missing. Such a model should map a complex
 core-shell ordering of the microwire.
 The inner core is composed of large ferromagnetic domains which are magnetized  
 parallel to the wire axis due to the dipole-induced shape anisotropy of the easy-axis type.
 In an external field these domains shrink or expand driving the DW propagation.
 Unlike the outer-shell observations, the experimental studies of the inner-core
 magnetization are indirect, thus, they require a model-based interpretation. 
 The minimal model for describing the remagnetization should include 
 the influence of a complex ordering of the outer shell (a bamboo-like
 patterned structure) on the inner-core DW.
 It would enable estimations of dynamical characteristics 
 with dependence on the parameters of the DW shape. An effective (analytical) approach 
 is preferable since, for typical (e.g. Fe-Si-B, Co-Si-B based)
 amorphous microwires, the presence of a strong internal stress in the wire 
 makes the simulations complex \cite{bet11,sto12}. More importantly, the (long-range)
 dipole (magnetostatic) spin interactions 
 interactions cannot be efficiently included into the numerical treatment since systems
 to consider are huge compared to the magnetostatic exchange length of the materials 
 (a parameter that limits the sizes of the wire discretization cell). 
 Note that the magnetostatics has been established to influence the DW-assisted
 remagnetization of the amorphous glass-coated microwires as well as the volume
 remagnetization of short (non-bistable) GMI microwires \cite{ye13,chi11,ipa14}.

The aim of the present paper is to provide the relevant model of the DW.
 The description of the inner core of the soft-ferromagnetic wire is formulated 
 within the Landau-Lifshitz-Gilbert (LLG) equation in 3D that we solve analytically.
 In the glass-coated amorphous or nanocrystalline ferromagnetic wires, the 
 magnetic ordering of the outer shell is due to a radially-distributed internal
 stress \cite{chi96,pha08,chi11a}. The stress is induced during the manufacturing
 process that contains the rapid solidification and slow cooling stages \cite{chi95,vel96,ant00}.
 The distribution of the stress can be then modified removing the glass coating or applying a thermal
 treatment \cite{chi97,ova09,chi13,kle14a}. Since the outer-shell pattern can be very complex and
 dependent on small fluctuations of the parameters of the manufacturing process, 
 we include the effect of that pattern on the inner core "fenomenologically", 
 (with a single numerical parameter only), via a Dzyaloshinskii-Moriya-like (DM) contribution
 to the LLG equation. 

A similar to the above described core-shell magnetic ordering of a ferromagnetic wire 
 can be induced without
 the internal stress while via a modulation of the wire diameter, thus, 
 the modulation of the magnetic-charge distribution on the wire surface \cite{cha12}.
 The present model of the inner-core
 ordering is applicable to such structured wires, which enables us to compare the analytical
 DW solutions to the model with the micromagnetically simulated DWs in the periodically
 constricted nanowires. 
 
Upon formulating the fundamentals of the model in Sec. II, in Sec. III, we present
 details of single-DW solutions. For a comparison, we perform micromagnetic simulations
 of the DWs in periodically constricted nanowires. In Sec. IV, we discuss aspects
 of the field-driven propagation of the DW which are connected to the model. 
 Conclusions are collected in Sec. V.
 
\section{Model}

Describing the inner core of a ferromagnetic wire, 
we consider the LLG equation in 3D in the form
\begin{eqnarray}
-\frac{\partial{\bf m}}{\partial t}=\frac{J}{M}{\bf m}\times\Delta{\bf m}
+\frac{\beta}{M}({\bf m}\cdot\hat{i}){\bf m}\times\hat{i}
+\gamma{\bf m}\times{\bf H}
\nonumber\\
-\frac{d}{M}{\bf m}\times\left(\hat{i}\times\frac{\partial{\bf m}}{\partial x}\right)
-\frac{\alpha}{M}{\bf m}\times\frac{\partial{\bf m}}{\partial t}.
\label{LLG}
\end{eqnarray}
Here, $\hat{i}\equiv(1,0,0)$, (the wire is directed along the $x$ axis), $M=|{\bf m}|$, $\gamma$ denotes the gyromagnetic factor,
 $J$ denotes the exchange constant, ${\bf H}=(H_{x},0,0)$ represents the external (longitudinal)
 magnetic field. The parameters $\beta$ and $d$ determine the strength of the effective 
 easy-axis anisotropy (of the magnetostatic origin, $\beta\le\gamma\mu_{0}M^{2}$) and a DM-like
 anisotropy, respectively.
 The DW solutions satisfy the condition ${\rm lim}_{|x|\to\infty}{\bf m}=\pm(M,0,0)$. 

The effective anisotropy of the DM type matches the properties of the dependence 
 on the magnetization derivative over the wire-axis coordinate (a consequence of the shell
 patterning) and of the non-invariance with respect to the space inversion. 
 The non-invariance results from the chirality of the ordering inside the outer shell.

In the periodically constricted wire, the DM-like anisotropy is due to a modulation
 of the projection of the wire-surface normal onto the long axis of the system.
 The inhomogeneous (periodic) distribution of the surface magnetic charges induce 
 a modulation of the outer-shell ordering, thus, modifying the axial distribution of the anisotropy
 in the inner core. In the amorphous glass-coated microwire, the origin of the outer-shell ordering 
 is different while the result of the mutual core and shell interaction is a similar helical
 ordering of the overall system \cite{cha12,chi07}. The outer-shell magnetization
 of the glass-coated amorphous microwire is circumferential in the case
 of the negative-magnetostriction materials while radial in the case of the positive
 magnetostriction. Additionally, the shell is divided into micro-domains because
 of the magnetostatics.

Looking for the solution to (\ref{LLG}), following \cite{bog80}, we apply the transform 
\begin{eqnarray}
m_{+}=\frac{2M}{f^{*}/g+g^{*}/f},
\hspace*{2em}
m_{x}=M\frac{f^{*}/g-g^{*}/f}{f^{*}/g+g^{*}/f},
\label{transform}
\end{eqnarray}
where $m_{\pm}\equiv m_{y}\pm{\rm i}m_{z}$. Thus, we introduce secondary dynamical parameters
 $g(x,y,z,t)$, $f(x,y,z,t)$ which take complex values and we ensure that the constraint $|{\bf m}|=M$
 is satisfied. The purpose is to obtain 
 the secondary equations of (unconstrained) motion in a tri-linear form, namely
\begin{eqnarray}
-f{\rm i}D_{t}f^{*}\cdot g=f\left[\alpha D_{t}+J(D_{x}^{2}+D_{y}^{2}+D_{z}^{2})
\hspace*{2em}\right.\nonumber\\\left.
+{\rm i}dD_{x}\right]f^{*}\cdot g
+Jg^{*}(D_{x}^{2}+D_{y}^{2}+D_{z}^{2})g\cdot g
\nonumber\\
-\left(\gamma H_{x}+\beta\right)|f|^{2}g,
\nonumber\\
-g^{*}{\rm i}D_{t}f^{*}\cdot g=g^{*}\left[\alpha D_{t}-J(D_{x}^{2}+D_{y}^{2}+D_{z}^{2})
\hspace*{2em}\right.\nonumber\\\left.
-{\rm i}dD_{x}\right]f^{*}\cdot g
-Jf(D_{x}^{2}+D_{y}^{2}+D_{z}^{2})f^{*}\cdot f^{*}
\nonumber\\
+\left(-\gamma H_{x}+\beta\right)|g|^{2}f^{*}.
\label{secondary-eq}
\end{eqnarray}
Here $D_{t}$, $D_{x}$, $D_{y}$, $D_{z}$ denote Hirota operators of differentiation 
$D_{x}^{n}b(x,y,z,t)\cdot c(x,y,z,t)\equiv
(\partial/\partial x-\partial/\partial x^{'})^{n}b(x,y,z,t)c(x^{'},y^{'},z^{'},t^{'})|_{x=x^{'},y=y^{'},z=z^{'},t=t^{'}}$.
Specific solutions of the multi-linear partial differential equations in the form 
of expansions over trial (exponential) functions correspond to multi-soliton solutions
to primary nonlinear equations. This fact enables a systematic study of multi-DW 
solutions to the LLG equation. In terms of the single-DW solutions, using (\ref{secondary-eq})
instead of (\ref{LLG}) is convenient when verifying the results of the next section.

\section{Domain-wall structure}

For the case $H_{x}=0$, Single-DW solution to (\ref{secondary-eq}) has been found in the form
\begin{eqnarray}
f=1,\hspace*{2em}g=u{\rm e}^{kx+qy+pz},
\label{solution_1}
\end{eqnarray}
where 
\begin{eqnarray} 
k^{2}+q^{2}+p^{2}-\frac{{\rm i}dk}{J}=\frac{\beta}{J}
\label{condition_1}
\end{eqnarray}
and ${\rm Re}k\neq 0$. We denote $k\equiv k^{'}+{\rm i}k^{''}$, $q\equiv q^{'}+{\rm i}q^{''}$, $p\equiv p^{'}+{\rm i}p^{''}$, 
 where $k^{'('')}$, $q^{'('')}$, $p^{'('')}$ take real values, rewriting (\ref{condition_1}) with 
\begin{IEEEeqnarray}{rCl}
k^{'2}+q^{'2}+p^{'2}-k^{''2}-q^{''2}-p^{''2}=\frac{\beta}{J}-\frac{dk^{''}}{J},
\label{condition_1_prime_a}\IEEEyessubnumber\\
k^{'}k^{''}+q^{'}q^{''}+p^{'}p^{''}-\frac{dk^{'}}{2J}=0.
\label{condition_1_prime_b}\IEEEyessubnumber
\end{IEEEeqnarray}
Assuming the DW to be centered at $(x,y,z)=0$, (then $u={\rm e}^{{\rm i}\phi}$), the relevant magnetization
 profile [the single-DW solution to (\ref{LLG})] is written explicitly with 
\begin{IEEEeqnarray}{rCl}
m_{+}(x,y,z)&=&M{\rm e}^{{\rm i}(\phi+k^{''}x+q^{''}y+p^{''}z)}{\rm sech}[k^{'}x+q^{'}y+p^{'}z],\nonumber\\
\label{profile1a}\IEEEyessubnumber\\
m_{x}(x,y,z)&=&-M{\rm tanh}[k^{'}x+q^{'}y+p^{'}z].
\label{profile1b}\IEEEyessubnumber
\end{IEEEeqnarray}
For $q=p=0$, the twisted transverse DW solution is obtained unless 
\begin{eqnarray}
k^{''}=\frac{d}{2J},\label{d-conditions_a}\\
k^{'2}=\frac{\beta}{J}-\frac{d^{2}}{4J^{2}}.\label{d-conditions_b}
\end{eqnarray}
For $q^{'}=p^{'}=0$ and $|k^{''}|\ll|q^{''}|=|p^{''}|,|k^{'}|$, the vortex DW of Fig. 1a
 is found unless the condition (\ref{d-conditions_a}) and 
\begin{eqnarray}
k^{'2}-2q^{''2}=\frac{\beta}{J}-\frac{d^{2}}{4J^{2}}
\label{nine}
\end{eqnarray}
are satisfied.
The solution of the type of the twisted transverse DW is expected not to be relevant to thick 
 wires since it corresponds to a magnetization that is constant within the cross sections
 of the inner core.

\begin{figure} 
\unitlength 1mm
\begin{center}
\begin{picture}(175,104)
\put(0,-6){\resizebox{88mm}{!}{\includegraphics{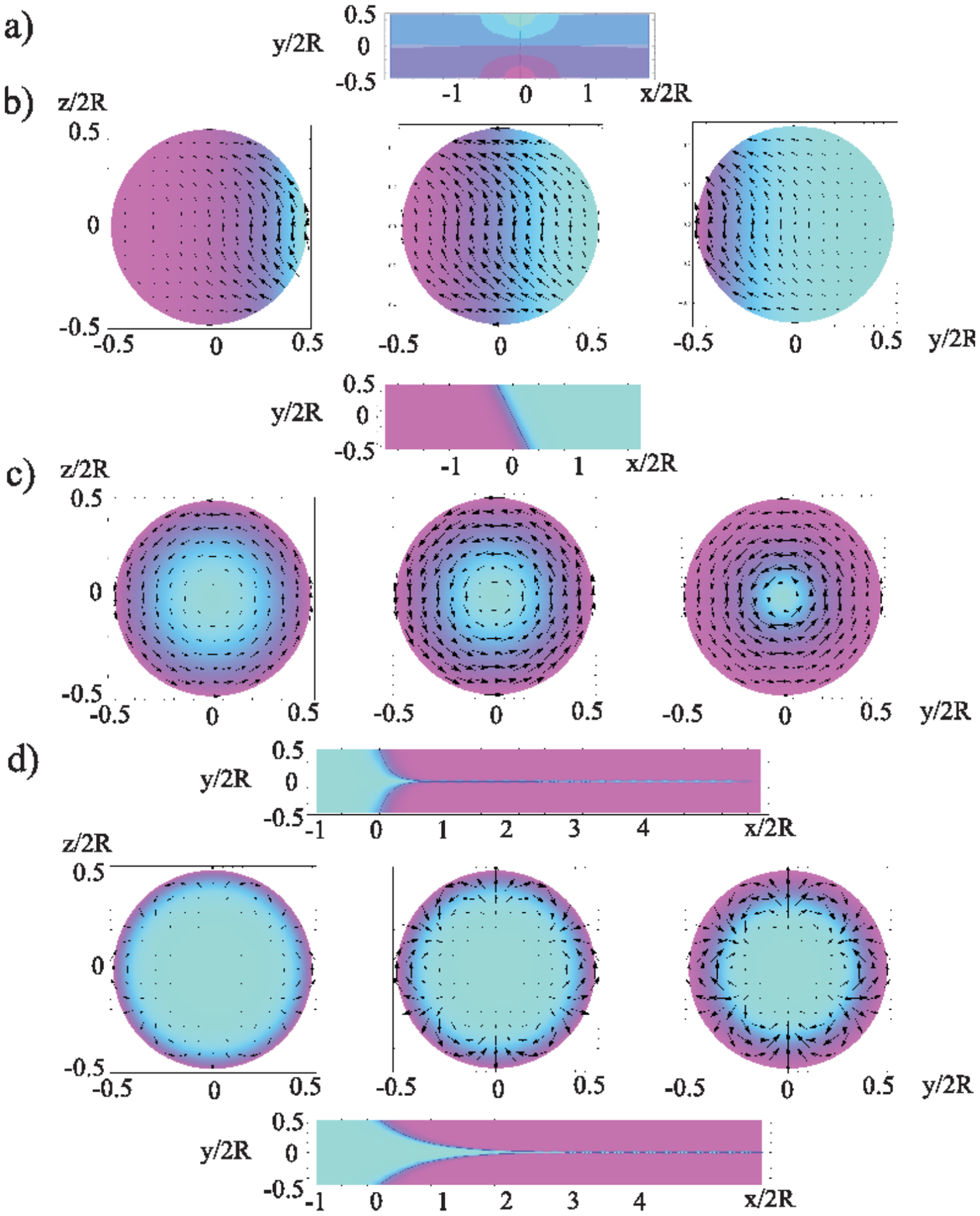}}}
\end{picture}
\end{center}
\caption{The magnetization in the cross-sections of the inner core in the vortex DW at $z=0$ (a), in the planar-DW at
$x=-q^{'}R/k^{'}$, $x=0$, $x=q^{'}R/k^{'}$, from left to right, and at $z=0$ (b), 
as well as the conical-DW at $x=-1/|k^{'}|$, $x=0$, $x=1/|k^{'}|$, from left to right, and at $z=0$,
for $\eta=1$ (c), and $\eta=4$ (d). For a)-d); $|k^{''}|\ll|k^{'}|=\pi/R$, $q^{'}/k^{'}=1/2$.
In a), the colors (red or blue) and its intensity indicate the sign and value of the z component
of the magnetization, whereas in b)-d), the sign and value of the x component of the magnetization. The arrows
indicate the projection of the magnetization onto the YZ plane.}
\end{figure}

Without loose of generality, we assume $p^{'}=0$, thus, we choose a DW orientation in the YZ plane. 
 The (dipole-induced) shape anisotropy that points the inner-core magnetization parallel to the 
 wire axis, drives the core-shell interface to be a local easy plane of the magnetization.
 We utilize this notice in order to reduce the number of free parameters.
 Namely, on the contour of the intersection of the  DW-plane ($k^{'}x+q^{'}y=0$)
 with the core-shell interface ($\sqrt{y^{2}+z^{2}}=R$), the magnetization should be aligned 
 in the circumferential direction wherever possible, in particular, at the central ($x=0$)
 and boundary ($x=\pm q^{'}R/k^{'}$) cross-sections of the wire in the DW-plane area.
 Therefore, $q^{''}=0$, $|\phi|=\pi/2$, $|p^{''}|R=\pi/2$. 
 Finally, from (\ref{condition_1_prime_a}), one finds (\ref{d-conditions_a}) and 
\begin{eqnarray}
k^{'2}+q^{'2}=\beta/J-d^{2}/4J^{2}+(\pi/2R)^{2}.
\label{ten}
\end{eqnarray}
to be satisfied. An additional "freedom" on the solution is due to independence of (\ref{LLG})
 of the  first derivatives of ${\bf m}$ 
 over $y$ and $z$ coordinates, which allows the perpendicular magnetization to take the form 
\begin{eqnarray} 
m_{+}(x,y,z)=M{\rm e}^{{\rm i}(\phi+k^{''}x+p^{''}|z|)}{\rm sech}[k^{'}x+q^{'}y] 
\end{eqnarray}
instead of (\ref{profile1a}). The relevant DW texture shown in Fig. 1b 
 corresponds to a "planar" DW of \cite{jim13}. A similar DW albeit deformed at its ends with
 a small flexture of the DW plane has been found with the micromagnetic simulations
 for the constricted nanowires \cite{cha12,bis77}. 

Using a different from (\ref{solution_1}) ansatz 
\begin{eqnarray}
f=1,\hspace*{2em}g={\rm e}^{kx+{\rm i}\phi+{\rm i}\eta\cdot{\rm arctan}(z/y)}\left(\frac{\sqrt{y^{2}+z^{2}}}{{\cal R}}\right)^{|\eta|}
\end{eqnarray}
whose map onto the YZ plane represents the (Belavin-Polyakov) Skyrmion \cite{jan12},
 we find it to satisfy (\ref{secondary-eq}) under the conditions (\ref{d-conditions_a})-(\ref{d-conditions_b}). We expect ${\cal R}$ to be close to $R$. 
 The relevant magnetization field is plotted in Figs. 1c, 1d, where the DW is seen 
 to be asymmetric relative to the YZ plane. 
 The present structure is singular at $(x,y,z)\to(\infty,0,0)$ (the DW is infinitely long 
 on the central line of the wire) and corresponds to a "tubular-conical" DW observed
 in \cite{moh90,pan12}. We expect the ordering in the DW to resemble the ordering in 
 the outer shell. Hence, the solution with $\eta=1$ is expected to be valid to the glass-coated 
 negative-magnetostriction wires, (in particular, Co-rich microwires), whose shell is circumferentially magnetized. The solutions with higher values of $|\eta|$ 
 are expected to describe the DWs in the wires of positive-magnetostriction materials,
 (in particular, Fe-rich microwires). Then $2(|\eta|-1)$ is 
 equal to the number of the radially-magnetized microdomains in each segment of the bamboo-structured outer shell. Obviously, the conical DW cannot be a stable 
 solution to the LLG equation. The elongated vertex of such a DW has to shrink with time 
in order to reduce the exchange energy. The instability of very elongated DWs in amorphous
 glass-coated microwires has been observed \cite{eks10}. A reported strong dependence 
 of the width of such DWs on the longitudinal external field coincides with the sensitivity 
 of the outer-shell ordering to the field and suggests the $\eta$ parameter of the DW to depend 
 on the field intensity \cite{gud09}. In the external field, the outer shell of 
 the negative-magnetostriction microwire can be of a complex vortex-containing texture,
 thus, we anticipate DWs of $|\eta|>1$ to be preferable in it \cite{chi04}.

\begin{figure} 
\unitlength 1mm
\begin{center}
\begin{picture}(175,84)
\put(0,-2){\resizebox{88mm}{!}{\includegraphics{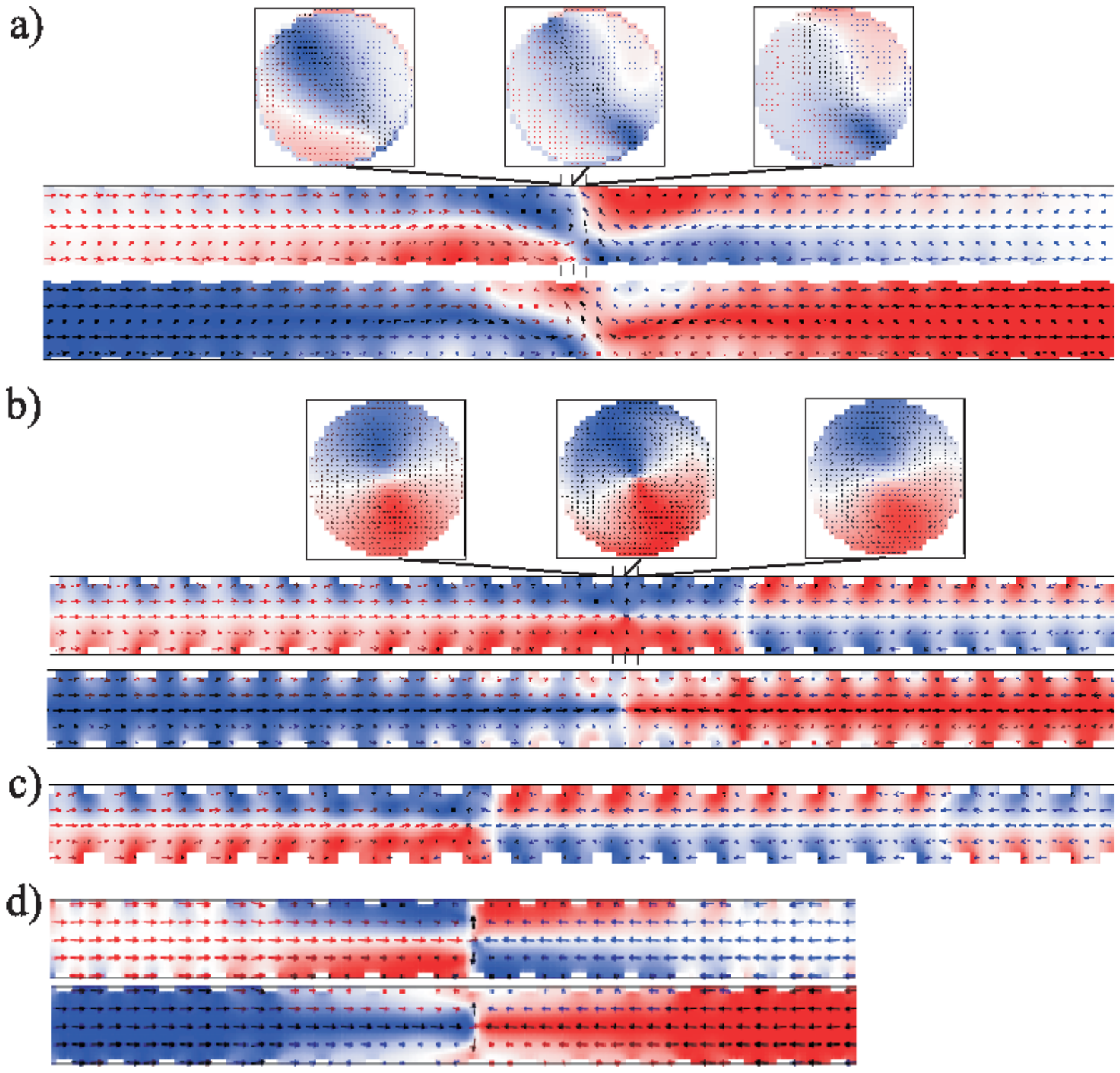}}}
\end{picture}
\end{center}
\caption{The magnetization in the longitudinal ($z=0$) and 
in the transverse ($x=$const) cross-sections 
of the constricted Py nanowire in the DW area obtained 
with the micromagnetic simulations.
The outer diameter of the wire is $D=150$nm, the constriction diameter is: $0.9D$ (a), $0.6D$ (b), $0.5D$ (c), 0.7D (d). 
Stable DWs of the planar type (a), vortex type (b), (c), and meta-stable (conical-like) DW (d)
are shown indicating the sign and value of the z (x) component 
of the magnetization at the top (bottom) pictures with the colors (red to blue).}
\end{figure}

In order to verify the validity of our effective approach in terms of predicting 
 the DW structure in core-shell ordered wires, we have performed micromagnetic
 simulations of the DWs in constricted soft-magnetic nanowires.
 We have considered Py nanowires (the saturation 
 magnetization $M=8.6\cdot10^{5}$A/m, the exchange stiffness $A_{ex}=1.3\cdot10^{-11}$J/m) 
 since Py is the most basic soft-magnetic material for the nanoscale applications.
 This makes our simulations realistic, while typical Metglas materials of the microwires
 are difficult to process to the nanoscale without inducing large internal stress. 
 Simulating infinitely-long nanowires, unlike \cite{cha12},
 we have discretized very (9$\mu$m) long systems. The discretization size was 5nm,
 (we utilized the OOMMF package), while the simulations were initialized with a stepwise
 magnetization distribution, (an infinitely narrow DW). The central 3$\mu$m-long fragment
 of the wire has been pre-calculated. Fixing the magnetization at the ends of the wire
 mesh, we counter the tendency towards reducing the surface magnetic charge. If we did not so,
 it would block the expansion of the bamboo-structure of the outer shell along the wire.
 We have kept constant the outer diameter $D=150$nm and the length of the constrictions while
 we have varied the diameter of the constrictions. Thus, we have controlled the strength
 of a helical anisotropy with a single parameter. 

According to Fig. 2, the radius of 
 the axially-magnetized area of the nanowire is found to be significantly smaller than
 the constriction radius, thus, the ordering is of a genuine core-shell type. The outer shell
 of the wire is seen to form a bamboo-like domain structure whose period decreases with decreasing
 the constriction diameter. The stable DWs are found to be of the planar type (Fig. 2a)
 or of the vortex type (Figs. 2b, 2c)
 for the constriction diameter bigger or smaller than about 0.8D, respectively.
 It is in a correspondence with our model that predicts the planar DWs to be formed
 provided $\beta>d^{2}/J$ because of (\ref{ten}), (a weak modulation of the surface layer
 of the wire), while the vortex-DW formation to allow the opposite relation $\beta<d^{2}/J$.
 The relaxation
 to the vortex DW structure as well as to the planar DW structure follows the formation
 of a meta-stable conical DW (Fig. 2d). In the simulations, we do not observe other DWs
 than predicted with our analytical model. Hence, we claim the model that combines
 the uniform easy-axis and a DM-like anisotropy to be capable to effectively describe
 the DWs of the core-shell ordered system. 

Let us notice that the vortex DWs have been previously found in the 
 circular-cross-section nanowires of the magnetocrystalline easy-plane anisotropy \cite{iva13},
 while the planar and meta-stable conical DWs have been predicted to result from
 the competition between the magnetostatics and complex exchange interactions \cite{bis77,oha75}.
 Our DW simulations enhance the motivation for describing both
 types of the DWs within a single model however.



With regard to the amorphous microwires, a simple estimation of $d$ is based on the disturbance of the stress-induced 
 magnetic field in the wire $H\sim 3|\lambda|\sigma/\mu_{0}M$, where $\lambda$ denotes 
 the magnetoelastic constant. 
 Here, the stress $\sigma=\sigma_{i}+\sigma_{a}$ consists of the internal and applied contributions.
 The internal stress is a complicated function 
 of the parameters of the wire composition and fabrication conditions (the initial temperatures of the wire and
 of the glass cover, an extraction stress) \cite{vel96}. 
 The first differential correction to this stress due to the ordering inhomogeneity (a magnetostatically-induced
 micro-domain structure) is expected to be of the order 
 of $\sigma l_{ms}|\nabla\times{\bf m}|/M$, where $l_{ms}=(2A_{ex}/\mu_{0}M^{2})^{1/2}$ denotes the magnetostatic exchange 
 length, ($A_{ex}=JM/2\gamma$). This leads to the estimate $d\sim \gamma3/2\cdot|\lambda|\sigma l_{ms}\equiv\gamma K_{me}l_{ms}$.

\section{Implications for domain-wall dynamics}
 
Let us denote any stationary single-DW solution to (\ref{secondary-eq}) by $f_{0}(x,y,z)$, $g_{0}(x,y,z)$. The inclusion 
 of $H_{x}\neq 0$ leads to the dynamical solutions $f(x,y,z,t)=f_{0}(x,y,z)=1$, $g(x,y,z,t)={\rm e}^{{\rm i}\omega t}g_{0}(x,y,x)$,
 where $\omega\equiv\omega^{'}+{\rm i}\omega^{''}=-\gamma H_{x}/(1+{\rm i}\alpha)$. It describes the driven motion
 of the DW with the velocity $v=\omega^{''}/k^{'}\equiv SH_{x}$ accompanied by the magnetization rotation about the wire axis 
 with the frequency $\omega^{'}$. Here, $S\approx\gamma\alpha/|k^{'}|$ denotes the DW mobility. However, the Walker breakdown
 is observed in the positive-magnetostriction amorphous and nanocrystalline microwires \cite{kle14}. In the regime
 of the viscous motion of the DW ($\omega^{'}=0$), taking the l.h.s. of (\ref{LLG}) 
 and (\ref{secondary-eq}) equal to zero, one finds $S=\gamma/\alpha|k^{'}|$. Since we consider the magnetization of the 
 outer shell of the wire to be stationary in the YZ plane, easy directions in the cross-sections of the inner core. Thus, the DW motion can be viscous \cite{chi09}. 
 For instance, the required discrete rotational symmetry of the shell is relevant to
 the conical DW of the parameter $|\eta|>1$.  

The measurements of the DW mobility for the glass-coated microwires with dependence on
 the applied stress has been performed in \cite{zhu12}.
 In the cited paper, a qualitative analysis for the viscous regime of the DW motion 
 has been performed taking $S=\gamma l_{me}/\alpha$. There, the width of the DW
 has been assumed to be the magnetoelastic exchange length $l_{me}=(2A_{ex}/K_{me})^{1/2}$.
 For sufficiently high stress applied, when 
 $k^{'2}\sim d^{2}/4J^{2}\gg\beta/J$, (the vortex-DW regime), the above assumption corresponds to taking $d\sim\gamma K_{me}l_{me}$ instead of 
 the previous-section estimate $d\sim\gamma K_{me}l_{ms}$. It is a consequence of neglecting the dipole-induced anisotropy
 while assuming the anisotropy in the outer shell of the wire to be of the purely magneto-elastic origin. In the consequence,
 the cited authors find $S\propto 1/\sqrt{\sigma}$, (in a disagreement with  experimental data), 
 while we evaluate $S=\gamma l_{me}^{2}/l_{ms}\alpha\propto 1/\sigma$.
 However, the magnetostatic origin of the domains inside the outer-shell has been established previously for
 the Fe-rich microwires \cite{ye13}, and for Co-rich microwires and nanowires \cite{chi11}.
 Moreover, following \cite{ant00}, the existence of the outer shell cannot be explained with
 the exchange and magnetoelastic spin interactions only, while the distribution 
 of the magnetic charges must be taken into account \cite{gud09}.
 A simple comparison of typical magnetoelastic and magnetostatic exchange lengths of the amorphous microwires
 ($K_{me}\sim10^{4}$J/m$^{3}$ for Fe-Si-B, $K_{me}\sim-10^{3}$J/m$^{3}$ for Co-Si-B, and $\mu_{0}M^{2}\sim10^{5}$J/m$^{3}$) 
 shows $l_{ms}^{2}/l_{me}^{2}\le1/10$ \cite{bet11,chi03}. Since the smallest characteristic exchange length
 is expected to determine the thickness of the outer-shell DWs, following the previous section, we claim the parameter $d$, 
 thus, the mobility of the inner-core DW to be dependent on the magnetostatic
 exchange length $l_{ms}$.

Despite the (Hirota) method that we apply is capable to treat the two-soliton states, 
 the description of the DW collision is complex because of the necessity of 
 including the dissipation. However, the result of
 the collision can be predicted on the basis of a simple rule that has been verified for ferromagnetic chains and stripes.
 When the magnetization in the closing up areas of the colliding DWs is parallel (antiparallel), the walls attract (repulse), 
 thus, they are expected to annihilate each other (to form a 2$\pi$-DW) \cite{jan13a,jan13b,jan13c}. 
 The circular (helical) ordering in the outer shell enforces the same direction of the magnetization
 curling in both the colliding DWs, (it drives the sign of the DW parameter $k^{''}$ to be the same
 as the sign of $d$). Therefore, we expect the collision to result in the annihilation of the DWs
 observed in \cite{zhu13a,zhu13b}.  

\section{Conclusions}

Within a simple model of the inner core of the amorphous ferromagnetic microwire, 
 we have established basic types of the DWs and determined the corresponding regimes
 of the LLG-equation parameters. The single constant of the effective DM-like
 anisotropy is capable to contain necessary information about the complex ordering
 of the outer shell that results from the permanent magnetostriction and magnetostatics.
 Correct evaluation of this constant allows interpretation of size and stress 
 effects in the DW motion. The model appeared to reproduce main features 
 of the DW shape in periodically-constricted nanowires. Hence, it passed
 a test of the validity to prototype systems which simulate the amorphous microwires,
 (while carrying out direct numerical simulations at the micrometer scale is an extremely
 challenging task). 

Finally, we mention the applicability of the effective DW description 
 to microwires coated with glassy and magnetic layers (bi-magnetic microwires) which are 
 of recent interest with regard to sensors \cite{tor13}.

\end{document}